\documentclass[a4paper]{article}
\usepackage[top=2.5cm,bottom=2.5cm,left=2.5cm,right=2.5cm]{geometry}
\usepackage{lmodern}
\usepackage[T1]{fontenc}

\DeclareFontShape{T1}{lmr}{bx}{sc}{<->ssub * lmr/bx/n}{}
\usepackage{graphicx}
\usepackage{xspace}
\usepackage{natbib}
\usepackage{booktabs}
\usepackage{array}
\usepackage{multirow}
\usepackage{amssymb}
\usepackage{enumitem}
\setlist{itemsep=2pt,topsep=3pt,parsep=0pt}
\usepackage{listings}
\usepackage{xcolor}
\lstdefinestyle{code}{
  basicstyle=\ttfamily\small,
  numbers=left,
  numberstyle=\tiny\color{gray},
  numbersep=6pt,
  breaklines=true,
  breakatwhitespace=false,
  breakindent=2em,
  columns=fullflexible,
  keepspaces=true,
  frame=single,
  framesep=4pt,
  xleftmargin=2em,
  xrightmargin=0.5em,
  aboveskip=6pt,
  belowskip=6pt,
}

\newcommand{\autospec}{\textsc{AutoSpec}\xspace}

\DeclareUnicodeCharacter{03BB}{\ensuremath{\lambda}}
\DeclareUnicodeCharacter{2190}{\ensuremath{\leftarrow}}
\DeclareUnicodeCharacter{2264}{\ensuremath{\leq}}
\DeclareUnicodeCharacter{2265}{\ensuremath{\geq}}
\DeclareUnicodeCharacter{211D}{\ensuremath{\mathbb{R}}}
\DeclareUnicodeCharacter{25B8}{\ensuremath{\triangleright}}
\DeclareUnicodeCharacter{2115}{\ensuremath{\mathbb{N}}}
\DeclareUnicodeCharacter{2223}{\ensuremath{\mid}}
\DeclareUnicodeCharacter{2200}{\ensuremath{\forall}}
\DeclareUnicodeCharacter{22A2}{\ensuremath{\vdash}}

\title{\autospec : Towards Verifiable Software Development via Automated Theorem Proving}
\author{
  Guchan Li\thanks{Department of Computer Science and Technology, Tsinghua University. Email: li-gc22@mails.tsinghua.edu.cn.} \and
  Hongning Wang\footnotemark[1]
}
\date{}

\begin{document}

\maketitle

\begin{abstract}
Formal verification offers guarantees beyond software testing, but brings significant technical challenges as trade off. We present \autospec{}, a Rust-to-Lean toolchain and benchmark that enables coding agents to prove mathematical specifications of existing program implementations. Our toolchain extends translation-based verification to numerical algorithms, by connecting program computation with Lean's mathematical abstraction. Our benchmark comprises 159 testcases across nine algorithmic domains and, to our knowledge, is the first to decouple implementation from specification proving and evaluate strong interactive coding agents. We further provide the first systematic analysis of proof structure and agent behaviour in this setting, identifying structural reasoning and verification-specific knowledge acquisition as key challenges. Based on these findings, we provide valuable insights suggesting concrete directions for future vericoding systems.
\end{abstract}

\section{Introduction}
Coding agents are rapidly changing how software is developed, making it increasingly important to establish the correctness of generated and modified programs. Testing can expose bugs, but cannot generally rule them out. Formal verification offers stronger guarantees through machine-checkable proofs, yet applying it to practical software remains challenging. General-purpose verification tools can restrict the expressiveness of specifications, while interactive theorem provers such as Lean offer a richer foundation for verification but traditionally require developers to implement programs in unfamiliar languages.

Translation-based verification offers a way to bridge this gap: programs are developed in a general-purpose language and mechanically translated into a representation suitable for formal reasoning. Aeneas provides such a workflow for Rust, translating programs into pure functional Lean code while exploiting Rust's ownership discipline to handle mutable references. This creates an opportunity to combine ordinary software development with the expressive specifications and mathematical libraries of Lean.

In this paper, we present \autospec{}, a pipeline and benchmark for studying automated verification of existing Rust implementations. We extend the Aeneas-based workflow with numerical abstraction and term simplification. In particular, numerical retargeting represents selected integer and floating-point computations using mathematical structures, allowing agents to reason about numerical algorithms with Mathlib. On top of this toolchain, we construct a benchmark of 159 programs across nine algorithmic domains, each equipped with a manually written specification. We evaluate eight coding agents in interactive Lean environments, asking them to prove the specifications of translated implementations.

We further go beyond aggregate proof accuracy to investigate the structure of generated proofs and the behavior of agents. Our analysis reveals substantial proof effort arising from loops and recursion, as well as knowledge acquisition challenge in navigating Aeneas and its supporting libraries. We further conducted experiments to address the latter problem. To our knowledge, this is the first study in this domain to systematically examine both vericoding proof structure and agent behavior, providing valuable insights into the bottlenecks that future systems must address. Our contributions are:

\begin{itemize}
    \item \textbf{Connecting software verification with mathematical reasoning.}
    We extend the Aeneas-based Rust-to-Lean workflow with numerical retargeting and term simplification. By representing selected numerical computations using mathematical structures, our toolchain enables agents to reason about translated software using Lean's mathematical definitions and Mathlib libraries.

    \item \textbf{Decoupling implementation from specification proving.}
    We introduce 159 testcases across nine algorithmic domains and two difficulty tiers. \autospec{} supplies existing Rust implementations and their mathematical specifications, isolating the task of proving that the implementations satisfy those specifications. To our knowledge, it is also the first benchmark in this domain to evaluate strong, interactive coding agents rather than chat models alone.

    \item \textbf{First analysis of vericoding proof structure and agent behaviour.}
    We characterize the structure of generated proofs and how coding agents interact with the verification environment. Our findings identify structural reasoning and verification-specific knowledge acquisition as important bottlenecks, and show that case-agnostic guidance can substantially improve proof success. These insights offer concrete directions for developing more capable vericoding systems.
\end{itemize}

\section{Background and Related Work}\label{sec:background}

We position \autospec{} among existing program verification approaches and benchmarks for vericoding.

\subsection{Software Verification Approaches}

Formal verification of general-purpose language (GPL) programs has a long history. Existing approaches can be broadly divided into three categories.

\paragraph{SMT-based verification.}
These approaches express program specifications through preconditions, postconditions, and invariants, and discharge the resulting proof obligations using automated solvers. Representative systems include Dafny~\citep{dafnylanguage}, Verus~\citep{verus}, Creusot~\citep{creusot}, Prusti~\citep{prusti}, and Kani~\citep{kani}. They provide relatively automated verification workflows, but their capabilities are constrained by the supported specification logic and underlying solving procedures.

\paragraph{Interactive theorem proving.}
Proof assistants such as Lean~\citep{lean}, Rocq~\citep{rocq}, and Isabelle~\citep{isabelle} support expressive specifications through dependent type systems and higher-order logic. However, developing verified programs in these languages requires substantial expertise and manual effort. Recent LLM-based agents have made significant progress in automated theorem proving~\citep{kimina, goedelv2, deepseekv2, compiletocompress, aristotle, hilbert, goedelarchitect, axproverbase, chen2025seedproverdeepbroadreasoning}, but have focused primarily on mathematical formalization rather than verifying existing GPL implementations.

\paragraph{Translation-based verification.}
An alternative is to separate implementation from verification by translating GPL programs into a verification language (VL). Tools such as hax~\citep{hax} and Aeneas~\citep{aeneas} support this approach for Rust, translating programs into functional representations suitable for theorem proving. In particular, Aeneas uses Charon~\citep{charon} and Rust's ownership discipline to translate mutable programs without explicitly modelling aliased heap state. \autospec{} builds on this approach to study automated verification of translated Rust implementations.

\subsection{Vericoding Benchmarks}

Existing benchmarks cover different stages of vericoding. Some evaluate the generation of specifications and proof annotations for existing programs~\citep{clover, mbpp-dafny, dafnybench, verusage}. Others study end-to-end verified code generation, asking models to produce specifications, implementations, and proofs~\citep{fvapps, cleverbench, verinabench}. Recent works have expanded evaluation across multiple verification languages, covering thousands of specifications and a range of classical algorithms~\citep{vericodingbench, algoveribench} 

Despite this progress, existing benchmarks do not systematically evaluate agents on real-world GPL implementations through mechanically generated VL representations. Some employ LLM-based translation, but such translations may introduce semantic drift and do not provide the same connection to the original implementation~\citep{fvapps, alphaverus, vericodingbench}. \autospec{} addresses this gap by deriving Lean programs from Rust implementations through an Aeneas-based pipeline, making mechanical translation an integral part of the verification workflow.

We choose Rust as the GPL and Lean 4 as the VL. Rust's ownership and borrowing system enables Aeneas to translate mutable programs without explicitly modeling aliased heap state~\citep{rustbelt}, while its use in systems and cryptographic software provides a source of practical algorithms. Lean provides access to Mathlib, a large library of formalized mathematics, allowing agents to reason about translated implementations using established mathematical definitions and theorems. It is also increasingly used in large-scale formalization and production software verification~\citep{sampcert}.

\section{\autospec Pipeline and Benchmark}

This section describes the Rust-to-Lean translation pipeline and the construction of the \autospec{} benchmark.

\subsection{Extended Aeneas Toolchain}\label{sec:toolchain}
The Rust compiler lowers source code to Mid-level Intermediate Representation (MIR), which Charon re-expresses as Low-Level Borrow Calculus (LLBC). LLBC makes ownership explicit: mutable references are represented as exclusive loans with associated back-functions that restore borrowed values when the loans end. Aeneas symbolically interprets LLBC and translates it into pure functional Lean definitions, typically expressed in the \texttt{Result} monad. By exploiting Rust's ownership discipline, Aeneas can track mutable loans without explicitly modeling aliased heap state.

Aeneas supports common scalar, slice, and vector operations, and its Lean backend integrates with Mathlib, which provides mathematical structures such as matrices and number systems. However, its numeric support limits its applicability to mathematical algorithms: floating-point arithmetic is unsupported, while fixed-width integers retain machine-level bounds and bit-vector semantics. We extend the toolchain in two ways: (1) numeric retargeting, which maps selected Rust numeric types to mathematical types in Lean, and (2) term simplification, which makes the generated definitions more amenable to mathematical reasoning.

\subsubsection{LLBC Retargeting}
Aeneas does not provide a floating-point model, preventing the direct translation of Rust programs containing \texttt{f64} operations. Rather than formalizing IEEE-754 arithmetic, we adopt mathematical numeric models for the algorithms in our benchmark. We implement this retargeting at the LLBC level by replacing selected numeric operations with calls to synthesized helper functions. These helpers are added to the LLBC program as a synthetic crate, allowing Aeneas to process them as ordinary function calls.

Numeric literals are handled similarly: each distinct constant is hoisted into a nullary helper function, avoiding the need for the target numeric type to support the original literal syntax. The adapter applies these transformations to statements, switch conditions, and tail expressions, while leaving crates without floating-point types unchanged.

Aeneas translates the modified LLBC program, after which a post-processing pass replaces helper calls with the corresponding operators and constants in Lean. This preserves the computational structure while changing the interpretation of the intercepted operations. Consequently, the resulting Lean definitions represent idealized mathematical computations rather than exact Rust floating-point executions. Specifications are written against these mathematical models, and proofs establish properties of the resulting retargeted programs.

\subsubsection{Numeric Models}
We select numeric models independently for integer and floating-point types in each testcase. Table~\ref{tab:models} summarizes the supported mappings.

For floating-point values, we use Lean's \texttt{Rat} or the real numbers $\mathbb{R}$. Rational arithmetic provides exact, executable operations, while real numbers support operations that are not generally closed over the rationals. For integers, we use \texttt{Int}, \texttt{Nat}, or fixed-width \texttt{BitVec}, depending on the intended abstraction. Signed integers map to \texttt{Int}; unsigned integers map to \texttt{Int} or \texttt{Nat}, unless fixed-width wrapping behaviour is required, in which case we use \texttt{BitVec}. We retain \texttt{usize} as \texttt{Std.Usize} for array and slice indexing.

\begin{table}[t]
    \centering
    \small
    \begin{tabular}{llll}
        \toprule
        Rust type & Bit-precise & Executable & Non-executable \\
        \midrule
        \texttt{f64} & --- & \texttt{Rat} & $\mathbb{R}$ \\
        \texttt{i8}--\texttt{i128}, \texttt{isize}
            & --- & \texttt{Int} & --- \\
        \texttt{u8}--\texttt{u128}
            & $\texttt{BitVec}\ w$ & \texttt{Int} / \texttt{Nat} & --- \\
        \texttt{usize}
            & \texttt{Std.Usize} & \texttt{Std.Usize} & --- \\
        \bottomrule
    \end{tabular}
    \caption{Numeric retargeting models, categorized by arithmetic semantics.}
    \label{tab:models}
\end{table}

These choices trade fidelity to machine-level arithmetic for proof tractability. Bit-precise models preserve fixed-width integer semantics, whereas executable and non-executable mathematical models abstract away machine-level bounds and floating-point rounding. Consequently, proofs under mathematical models establish properties of idealized computations rather than bit-precise properties of the original Rust executions. This abstraction allows us to focus the benchmark on the mathematical correctness of algorithms.

\subsubsection{Term Simplification}
Aeneas represents computations using the \texttt{Result} monad, which can make even simple arithmetic expressions cumbersome for mathematical reasoning. Our post-processing pass simplifies generated definitions by rewriting arithmetic over retargeted types into pure expressions. It also handles adapter-specific proof obligations, marks definitions that depend on axioms as non-computable, and restores a top-level module layout so that loop definitions can be evaluated by Lean's interpreter.

\subsubsection{Specification Style and Mathlib}
The mathematical numeric models allow specifications to express algorithmic properties directly. For example, Appendix~\ref{app:lu-example} presents a specification for LU decomposition with partial pivoting. We use Mathlib's abstractions to express properties such as finite sums and matrix operations. This lets agents reuse established mathematical definitions and lemmas rather than formalizing them from scratch. Their main task is to connect the translated computation to its specification, for example, by discovering suitable loop invariants and intermediate properties.

\subsection{Benchmark Overview}\label{sec:benchmark}

\begin{table}[t]
      \centering
      \small
      \setlength{\tabcolsep}{3.5pt}
      \begin{tabular}{lrrrrrrrrrr}
            \toprule
            Category        & array & crypto & dp  & ds  & graph & integer & matrix           & string & tensor       & \textbf{Overall} \\
            \midrule
            Cases           & 15    & 10     & 20  & 20  & 34    & 10      & 20               & 25     & 5            & \textbf{159}     \\
            Tier 1          & 7     & 5      & 12  & 10  & 17    & 10      & 10               & 17     & 5            & \textbf{93}      \\
            Tier 2          & 8     & 5      & 8   & 10  & 17    & 0       & 10               & 8      & 0            & \textbf{66}      \\
            \midrule
            Rust lines      & 68    & 109    & 94  & 316 & 211   & 30      & 186              & 219    & 40           & \textbf{171}     \\
            Lean func lines & 156   & 245    & 188 & 289 & 349   & 78      & 676              & 244    & 130          & \textbf{297}     \\
            Lean spec lines & 43    & 38     & 70  & 101 & 134   & 22      & 83               & 106    & 76           & \textbf{87}      \\
            \bottomrule
      \end{tabular}
      \caption{Benchmark composition.}
      \label{tab:benchmark}
\end{table}

The benchmark comprises 159 testcases across nine algorithmic domains, summarized in Table~\ref{tab:benchmark}. We divide them into two tiers according to code complexity. Tier 1 contains 93 relatively small, everyday algorithmic programs, including sorting, standard data structures, classical number theory, basic graph algorithms, and dynamic programming. They average 87 Rust lines and 155 Lean function lines per testcase, and their specifications typically express direct mathematical identities.

Tier 2 contains 66 larger programs, averaging 289 Rust lines and 497 Lean function lines per testcase. Many are drawn from competitive programming and algorithm repositories. Their specifications involve named theorems from the literature, requiring mathematical insights beyond what is immediately apparent from the program's control flow. The tier assignment is based on program size, with category-specific thresholds to account for differences in the natural scales of different algorithm domains.

\subsubsection{Construction Process}
For each testcase, we prepare a self-contained Rust function implementing the target algorithm. We normalize reference implementations to avoid Rust constructs unsupported by Charon or Aeneas, removing object-oriented structure and encapsulation where necessary to keep the verification task focused on the algorithm. We then translate the function using our extended Aeneas toolchain and manually write a Lean correctness theorem specifying its output in terms of standard mathematical objects from Mathlib.

The construction process illustrates a practical vericoding workflow: a coding agent implements a Rust function, a translation tool produces its Lean representation, and a human engineer provides and approves the formal specification. The benchmark instantiates this workflow across a collection of algorithms, with the resulting Lean proofs serving as the evaluation task for coding agents.

\subsubsection{Quality Controls}

We apply three checks to each testcase to assess implementation correctness, translation consistency, and specification quality: (1) \textbf{Rust tests.} We manually inspect the source and run its Rust unit tests on representative inputs. (2) \textbf{Translation consistency.} We translate Rust unit tests into executable Lean functions and use \texttt{\#eval} assertions to compare their outputs. (3) \textbf{Specification review.} A reviewer agent identifies edge cases that could otherwise make a theorem unprovable or trivially provable for unintended reasons.  Together, these checks provide complementary evidence for the reliability of the benchmark.

\section{Experiments}\label{sec:experiment}

We evaluate eight coding agents on our two-tier benchmark. Beyond proof accuracy, we investigate their problem-solving behavior and the structure of the proofs they generate.

\subsection{Accuracy across Categories}
We evaluate Claude Opus 5.5, Claude Opus 4.5, GPT-6 Sol, GPT-5.6 Sol, DeepSeek V4.1 Flash, GLM 5.3, Kimi K3, and Leanstral 1.5. A proof counts as successful only if it is accepted by Lean Comparator, which rejects proofs that alter the theorem statement or rely on non-standard axioms.

\begin{table}[htbp]
    \centering
    \caption{Evaluation results. $n$ is the number of testcases.}
    \label{tab:accuracy}
    \small
    \setlength{\tabcolsep}{3pt}
    \begin{tabular}{lrrrrrrrrr}
        \toprule
                           & n   & \rotatebox{60}{Opus 5.5} & \rotatebox{60}{Opus 4.5} & \rotatebox{60}{GPT 6} & \rotatebox{60}{GPT 5.6} & \rotatebox{60}{DS V4.1} & \rotatebox{60}{GLM 5.3} & \rotatebox{60}{Kimi K3} & \rotatebox{60}{Leanstral} \\
        \midrule
        array              & 15  & 14                       & 1                        & 3                     & 3                       & 2                       & 1                       & 2                       & 0                         \\
        crypto             & 10  & 9                        & 2                        & 2                     & 2                       & 4                       & 6                       & 3                       & 1                         \\
        dp                 & 20  & 14                       & 0                        & 2                     & 6                       & 0                       & 0                       & 0                       & 0                         \\
        ds                 & 20  & 16                       & 0                        & 8                     & 2                       & 3                       & 1                       & 0                       & 0                         \\
        graph              & 34  & 17                       & 0                        & 2                     & 0                       & 0                       & 0                       & 0                       & 0                         \\
        integer            & 10  & 10                       & 3                        & 3                     & 2                       & 8                       & 7                       & 5                       & 3                         \\
        matrix             & 20  & 12                       & 0                        & 0                     & 0                       & 0                       & 0                       & 0                       & 0                         \\
        string             & 25  & 19                       & 0                        & 7                     & 4                       & 0                       & 0                       & 0                       & 0                         \\
        tensor             & 5   & 5                        & 0                        & 3                     & 1                       & 0                       & 1                       & 0                       & 0                         \\
        \midrule
        Harness$^\dagger$            & -   & cc                       & cc                       & cx                    & cx                      & dsh                     & cc                      & cc                      & vb                        \\
        Max turns          & -   & 100                      & 100                      & 100                   & 100                     & 100                     & 100                     & 100                     & 400                       \\
        Cost (\$) &  -   & 12.3                     & 31.6                     & 8.7                   & 5.6                     & 0.35                    & 8.3                     & 15.7                    & $\approx 0$               \\
        \midrule
        Total (\%)         & 159 & 73.0                     & 3.8                      & 18.9                  & 12.6                    & 10.7                    & 10.1                    & 6.3                     & 2.5                       \\
        Tier 1 (\%)        & 93  & 89.2                     & 6.5                      & 25.8                  & 17.2                    & 16.1                    & 16.1                    & 10.8                    & 4.3                       \\
        Tier 2 (\%)        & 66  & 50.0                     & 0.0                      & 9.1                   & 6.1                     & 3.0                     & 1.5                     & 0.0                     & 0.0                       \\
        \bottomrule
    \end{tabular}
    \\[2pt]
    {\footnotesize $^\dagger$ cc = Claude Code, cx = Codex, dsh = DeepSeek Harness, vb = Mistral Vibe.}
\end{table}

Each agent runs in an isolated Lean workspace with access to a working LSP MCP server. We impose a budget of 100 tool calls per testcase, except for Leanstral 1.5 with a limit set to 400. We report accuracy as the percentage of testcases accepted by Lean Comparator. Table~\ref{tab:accuracy} summarizes the number of solved testcases and the aggregate costs. Leanstral 1.5 is deployed locally, so its cost is approximately zero.

Performance varies substantially across agents, and the benchmark remains far from saturated. Opus 5.5, solves 73.0\% of the testcases and less than a half on Tier 2. Opus 4.5 and Leanstral 1.5 achieve the lowest accuracy, primarily in the integer and cryptography categories. Graph, matrix, and DP are particularly challenging across agents. All eight agents perform worse on Tier 2 than on Tier 1, indicating that the benchmark's more complex tasks remain challenging even for strong models.

\begin{table}[htbp]
    \centering
    \caption{Accuracy of the best-performing model on each benchmark.}
    \label{tab:comparison}
    \small
    \setlength{\tabcolsep}{4pt}
    \begin{tabular}{lrllrr}
        \toprule
        Benchmark  & Tasks & Best model      & Protocol & Budget      & Accuracy         \\
        \midrule
        DafnyBench & 782   & Claude 3 Opus   & chat     & 10 attempts & 67.8\%$^\dagger$ \\
        CLEVER     & 161   & Claude 3.7      & chat     & 600\,s      & 8.69\%           \\
        VERINA     & 189   & GPT o3, o4-mini & chat     & 1 attempt   & 3.2\%            \\
        Vericoding & 2361  & GPT-5           & chat     & 5 attempts  & 17.9\%           \\
        AlgoVeri   & 77    & GPT-5.3 Codex   & mixed    & 8 rounds    & 11.7\%           \\
        \midrule
        \autospec  & 159   & Claude Opus 5.5 & harness  & 100 turns   & 73.0\%           \\
        \bottomrule
    \end{tabular}
    \\[2pt]
    {\footnotesize $^\dagger$ Dafny; all other numbers are for Lean.}
\end{table}

We place \autospec{} in the context of existing vericoding benchmarks (Table~\ref{tab:comparison}). Two aspects distinguish our evaluation:

\begin{itemize}
    \item \textbf{Agent-based evaluation.}
    Among the benchmarks listed, \autospec{} is distinguished by evaluating complete coding agents rather than only model responses in a prescribed interaction loop. Agents can inspect proof states, test tactics incrementally, read library source code, and recover from errors. This more closely reflects how developers interact with an interactive theorem prover.
    \item \textbf{Extended interaction budget and recent models.}
    Our agents receive up to 100 tool calls per testcase, compared with the 1--10 attempts reported by several other benchmarks. This allows substantially more interaction and feedback during proof development. We also evaluate more recent models, making it possible to assess whether the benchmark remains challenging under relatively favorable evaluation conditions.
\end{itemize}

By providing mechanically translated programs and formal specifications, our benchmark isolates the challenge of constructing machine-checkable proofs, enabling a focused assessment of agents' verification capabilities. Together with its interactive agent setting and generous tool budget, this design provides a valuable testbed for studying how coding agents reason about and verify real-world algorithms.

\subsection{Proof Structure Analysis}
Program verification differs from classical mathematical deduction due to its unique operational structures. To characterize this, we analyze the structures of successful proofs produced by Claude Opus 5.5, which has the highest number of passed cases. Proof tree is defined in Appendix~\ref{app:proof-tree} in detail.

\subsubsection{Individual Theorem}

\begin{figure}[htbp]
    \centering
    \includegraphics[width=0.8\linewidth]{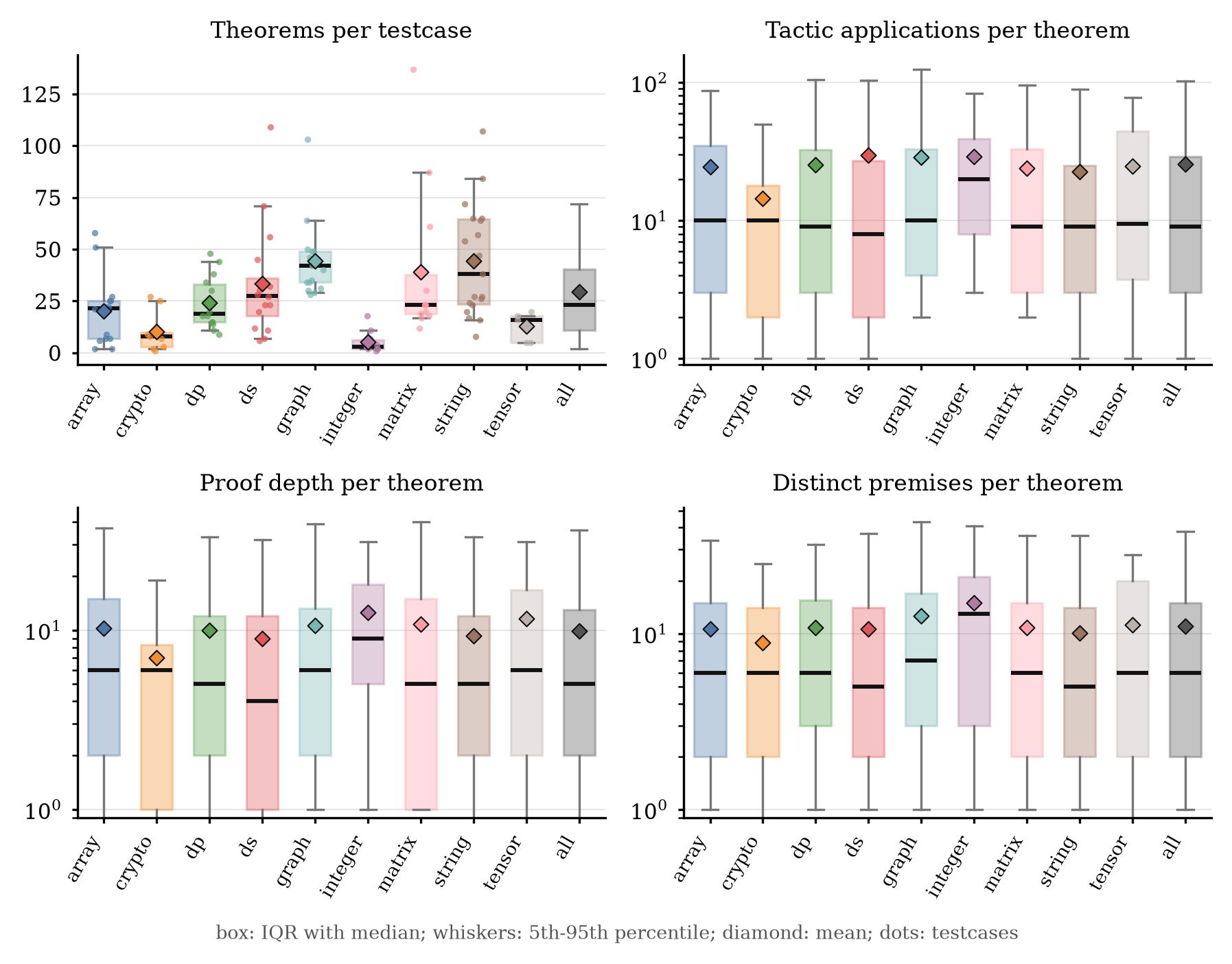}
    \caption{Proof-tree size by benchmark category.}
    \label{fig:proof_size}
\end{figure}

We measure proof size along three complementary dimensions. \textbf{Tactics} counts all tactic applications in the proof tree, measuring the total amount of proof construction. \textbf{Depth} measures the longest sequence of dependent reasoning steps from the root to a leaf. \textbf{Premises} counts the distinct cited lemmas, serving as a proxy for the breadth of knowledge required to construct the proof. Figure~\ref{fig:proof_size} reports the distributions of these metrics across benchmark categories, together with the number of theorems per testcase.

The size of individual theorems is similar across most categories: the median theorem requires 8--10 tactic applications in every category except integer. However, the distribution is highly skewed: the largest 10\% of theorems account for 48.5\% of all tactic applications. Based on their names, we find that these disproportionately large theorems are primarily associated with loops, recursion, and invariants.

\subsubsection{Theorem organization}
Beyond individual theorem size, proof complexity depends on how theorems are organized and combined. Each testcase contains 30 theorems on average, ranging from 5 in integer to 44 in graph and string. To characterize the dependencies among these obligations, we construct a theorem dependency graph for each testcase. Each node represents a theorem, and a directed edge connects a theorem to the lemmas it cites as premises.

\begin{table}[htbp]
    \centering
    \caption{Theorem dependency graph statistics.}
    \label{tab:dependency}
    \small
    \setlength{\tabcolsep}{3.5pt}
    \begin{tabular}{lrrrrrrrrrr}
        \toprule
                                   & array & crypto & dp   & ds   & graph & integer & matrix & string & tensor & \textbf{Total} \\
        \midrule
        Longest chain              & 4.7   & 3.0    & 5.9  & 5.4  & 9.1   & 3.1     & 6.5    & 7.5    & 5.4    & \textbf{6.0}   \\
        Widest layer               & 10.6  & 6.1    & 13.1 & 16.0 & 20.9  & 2.7     & 22.2   & 21.6   & 6.8    & \textbf{15.0}  \\
        Critical path share (\%)   & 58    & 62     & 57   & 45   & 44    & 89      & 43     & 47     & 86     & \textbf{55}    \\
        \bottomrule
    \end{tabular}
\end{table}

Table~\ref{tab:dependency} summarizes the resulting graph structures. The longest chain measures the maximum dependency depth, while the widest layer counts the maximum number of mutually independent theorems at the same level. The critical path share measures the fraction of all tactic applications contained in the theorems along the longest dependency chain, and 100\% indicates a fully linear proof.

The results reveal two contrasting structures. \textbf{Wide proofs}: graph, string, matrix, and data-structure testcases contain many theorems, but only 43--47\% of their tactic applications lie on the critical path. Their proofs distribute the work across many relatively independent lemmas, which are subsequently combined. \textbf{Linear proofs}: integer and tensor testcases have fewer theorems, with 86--89\% of tactic applications on the critical path. Their proofs are dominated by a sequence of dependent deductions, often concentrated in a few deeper theorems.

These results show that similar-sized individual theorems can be assembled into very different proof structures. Overall, proof complexity is shaped not only by the size of individual reasoning steps, but also by how the program induces and organizes its proof obligations.

\subsubsection{Tactic Composition}
We further characterize proof construction by grouping tactic applications into eight functional categories, summarized in Table~\ref{tab:tactic_buckets}. To provide a reference point for conventional mathematical theorem proving, we also analyze the same tactic categories in a Lean formalization of the Navier--Stokes and Euler blow-up results, a research-level development written by GPT-6 Astra.

\begin{table}[htbp]
    \centering
    \caption{The eight tactic buckets.}
    \label{tab:tactic_buckets}
    \small
    \begin{tabular}{lll}
        \toprule
        Bucket & Role                                & Example                                                                          \\
        \midrule
        aen    & Aeneas symbolic execution           & \texttt{step}, \texttt{scalar\_tac}                                              \\
        math   & math decision procedures            & \texttt{omega}, \texttt{ring}, \texttt{linarith}, \texttt{decide}                \\
        simp   & rewriting by simplification         & \texttt{simp}, \texttt{simp\_all}, \texttt{aesop}, \texttt{gcongr}               \\
        rw     & explicit directed rewriting         & \texttt{rw}, \texttt{conv}, \texttt{symm}                                        \\
        dec    & goal decomposition and case splits  & \texttt{cases}, \texttt{rcases}, \texttt{obtain}, \texttt{induction}             \\
        ib     & hypothesis introduction and binding & \texttt{intro}, \texttt{have}, \texttt{let}, \texttt{set}, \texttt{unfold}       \\
        cl     & goal-closing steps                  & \texttt{exact}, \texttt{apply}, \texttt{refine}, \texttt{rfl}, \texttt{calc}     \\
        cmb    & tactic glue                         & \texttt{<;>}, \texttt{first}, \texttt{try}, \texttt{all\_goals}, \texttt{repeat} \\
        \bottomrule
    \end{tabular}
\end{table}

Across all benchmark categories, structural tactics account for a large share of proof construction. Goal decomposition, hypothesis introduction and binding, and explicit rewriting together make up 59.5\% of all tactic applications. In contrast, mathematical decision procedures account for 9.8\%. Graph proofs exhibit the highest rate of goal decomposition and the lowest use of mathematical automation, consistent with the need to reason about inductively defined graph predicates.

\begin{table}[htbp]
    \centering
    \caption{Tactic composition by category (\%). N--S is the Navier--Stokes reference.}
    \label{tab:tactic_composition}
    \small
    \setlength{\tabcolsep}{3.5pt}
    \begin{tabular}{lrrrrrrrrrr@{\hspace{12pt}}r}
        \toprule
             & array & crypto & dp   & ds   & graph & integer & matrix & string & tensor & \textbf{Total} & N--S \\
        \midrule
        aen  & 2.1   & 0.5    & 2.8  & 0.3  & 0.2   & 0.1     & 4.5    & 1.5    & 2.0    & \textbf{1.5}   & 0.0  \\
        math & 10.8  & 12.1   & 12.9 & 7.0  & 7.3   & 14.5    & 9.8    & 12.5   & 10.3   & \textbf{9.8}   & 12.1 \\
        simp & 15.7  & 21.8   & 14.1 & 27.6 & 12.9  & 13.7    & 12.5   & 13.6   & 12.6   & \textbf{16.1}  & 11.3 \\
        rw   & 12.1  & 13.7   & 11.2 & 7.9  & 11.5  & 13.1    & 11.6   & 12.2   & 14.5   & \textbf{11.2}  & 10.0 \\
        dec  & 26.5  & 27.1   & 25.9 & 28.7 & 30.6  & 22.9    & 23.5   & 27.2   & 26.9   & \textbf{27.6}  & 11.8 \\
        ib   & 21.4  & 14.7   & 21.0 & 15.2 & 21.8  & 24.8    & 25.1   & 21.1   & 21.5   & \textbf{20.7}  & 33.3 \\
        cl   & 9.8   & 6.2    & 9.9  & 9.1  & 14.4  & 9.7     & 11.7   & 10.9   & 9.2    & \textbf{11.2}  & 19.7 \\
        cmb  & 1.6   & 3.8    & 2.0  & 4.2  & 1.3   & 1.2     & 1.3    & 0.9    & 3.0    & \textbf{1.9}   & 1.8  \\
        \bottomrule
    \end{tabular}
\end{table}

Table~\ref{tab:tactic_composition} compares tactic composition across the benchmark and the Navier--Stokes reference corpus. Both use mathematical decision procedures at comparable rates (9.8\% and 12.1\%), but differ substantially in how they organize the remaining proof work. Vericoding proofs use goal decomposition more than twice as often (27.6\% versus 11.8\%), reflecting the need to handle branches and loop iterations in translated programs. The Navier--Stokes proofs instead devote more steps to introducing intermediate claims and local definitions (33.3\% versus 20.7\%), as well as closing goals (19.7\% versus 11.2\%).

This contrast highlights a structural difference between program verification and conventional mathematical formalization. Vericoding proofs are shaped by the control flow and representation of the implementation, whereas mathematical proofs are more often organized around chains of intermediate claims. Although mathematical reasoning remains essential, a substantial part of the vericoding challenge lies in decomposing program behavior and managing the resulting proof obligations.

\subsection{Agent Behavior Analysis}
We investigate how agents construct proofs, focusing on the knowledge they rely on and how they acquire it. We first analyze the provenance of cited lemmas to identify the knowledge sources used in successful proofs. We then examine agents' tool-call traces and file-reading behaviour, and test whether targeted guidance can improve their exploration efficiency.

\subsubsection{Premise Provenance}
We classify the premises (cited lemmas) in successful proofs according to where they are defined: (1) \textbf{Aeneas.} The Aeneas standard library provides infrastructure for reasoning about translated programs, including specialized tactics such as \texttt{step} for symbolic execution. (2) \textbf{Program.} The testcase's own module contains definitions and lemmas about the translated implementation. Reasoning about loops and recursion often requires establishing intermediate properties and invariants. (3) \textbf{Math.} Mathlib and the Lean core provide general mathematical definitions and theorems, supporting the mathematical specifications of the benchmark.

\begin{table}[htbp]
    \centering
    \caption{Premise provenance by category, as \% of premise citations.}
    \label{tab:premise_provenance}
    \small
    \setlength{\tabcolsep}{3.5pt}
    \begin{tabular}{lrrrrrrrrrr}
        \toprule
                & array & crypto & dp   & ds   & graph & integer & matrix & string & tensor & \textbf{Total} \\
        \midrule
        Aeneas  & 22.6  & 27.4   & 20.8 & 10.3 & 15.8  & 18.0    & 24.0   & 15.6   & 32.5   & \textbf{17.4}  \\
        Program & 27.7  & 23.9   & 27.7 & 42.0 & 28.5  & 13.7    & 31.8   & 31.0   & 21.1   & \textbf{31.2}  \\
        Math    & 49.7  & 48.7   & 51.4 & 47.7 & 55.7  & 68.4    & 44.1   & 53.4   & 46.5   & \textbf{51.3}  \\
        \bottomrule
    \end{tabular}
\end{table}

Table~\ref{tab:premise_provenance} reports the proportion of premise citations from each source. Overall, 51.3\% come from Mathlib or the Lean core, 31.2\% from the testcase's own module, and 17.4\% from Aeneas. The distribution varies across categories: integer proofs rely most heavily on mathematical results, consistent with their number-theoretic specifications. Tensor, cryptography, and matrix proofs make greater use of Aeneas infrastructure, reflecting the additional reasoning needed to handle translated computations involving arrays and arithmetic.

These results suggest that successful verification requires two complementary kinds of knowledge: mathematical knowledge for establishing the specification, and program-specific knowledge for reasoning about the translated implementation. We next investigate how agents acquire these resources through tool use.

\subsubsection{Reading Behaviour}
\label{sec:hmm}
We model each agent's tool-call trace as a sequence of actions, mapping each call to one of five symbols: \texttt{read}, \texttt{write}, \texttt{build}, \texttt{mcp}, or \texttt{other}. We fit a categorical hidden Markov model (HMM) to the pooled traces of all eight agents using Baum--Welch~\citep{baum-welch}. The shared model assigns the same meaning to each hidden state across agents. We use five hidden states and name each phase after its most frequent emitted symbol. Appendix~\ref{app:hmm} reports the emission matrix and further analyzes agent-specific transition patterns.

\begin{table}[htbp]
    \centering
    \caption{Share of each agent's tool calls decoded into each phase (\%).}
    \label{tab:hmm_occupancy}
    \begin{tabular}{lrrrrr}
        \toprule
        Agent     & read & write & write+read & mcp  & mcp+read \\
        \midrule
        Opus 5.5  & 16.4 & 26.8  & 18.7       & 2.8  & 35.3     \\
        Opus 4.5  & 25.6 & 15.5  & 15.5       & 18.6 & 24.7     \\
        \addlinespace
        DS V4.1   & 38.0 & 6.0   & 45.6       & 2.9  & 7.5      \\
        Kimi K3   & 48.3 & 13.0  & 13.4       & 10.0 & 15.4     \\
        GLM 5.3   & 53.0 & 4.2   & 20.8       & 16.9 & 5.1      \\
        \addlinespace
        GPT 5.6   & 27.2 & 0.1   & 72.7       & 0.0  & 0.0      \\
        GPT 6     & 23.5 & 0.8   & 72.5       & 2.2  & 1.0      \\
        \addlinespace
        Leanstral & 35.3 & 1.9   & 5.8        & 54.0 & 2.9      \\
        \bottomrule
    \end{tabular}
\end{table}

Agents differ substantially in their tool-use patterns (Table~\ref{tab:hmm_occupancy}), particularly in how much they read. The read phase accounts for 16\% of Opus 5.5's calls, compared with 53\% for GLM 5.3. The agents also exhibit distinct workflows. The two Claude models distribute much of their remaining activity between drafting and language-server queries. The GPT agents concentrate their drafting in the write+read phase and rarely query the language server. Leanstral 1.5 takes the opposite approach, spending more than half of its calls in the MCP phase.

Reading also serves different roles across agents. After leaving the read phase, Opus 5.5, DeepSeek V4.1, and GPT-6 transition to drafting in 97\%, 87\%, and 98\% of cases, respectively. For GLM 5.3 and Leanstral 1.5, these rates are only 43\% and 18\%, with reading more often followed by language-server interactions. Thus, agents differ not only in how much they read, but also in how they incorporate retrieved information into their workflows.

\subsubsection{File Read Analysis}
\label{sec:file-read}
The HMM analysis treats all reads alike. We therefore examine which library files agents access and whether their final proofs use those files. For each successfully proved testcase, we record three sets:

\begin{itemize}
    \item $\mathit{DR}$ (direct reads): files explicitly opened through the file-reading tool.
    \item $\mathit{IR}$ (indirect reads): files encountered through other tool calls, such as Lean elaboration, without being explicitly opened.
    \item $C$ (citations): files containing lemmas cited in the final proof. We exclude lemmas resolved implicitly by automation, such as \texttt{omega}.
\end{itemize}

Let $R = DR \cup IR$ denote all accessed files. We define two ratios:

\[
    \eta = \frac{|R \cap C|}{|R|},
    \qquad
    \rho = \frac{|R \cap C|}{|C|}.
\]

The \emph{exploration efficiency} $\eta$ measures the fraction of accessed files that contribute cited lemmas; a low value indicates that much of the exploration does not appear in the final proof. The \emph{pretraining coverage} $\rho$ measures the fraction of cited files that the agent accessed during the run. A low value suggests that the agent relied on prior knowledge rather than in-context exploration for many of its cited lemmas. Table~\ref{tab:file_read} reports these metrics by agent and library. We exclude Opus 4.5 and Leanstral 1.5 because they have too few successful cases for this analysis.

\begin{table}[htbp]
    \centering
    \caption{Read/use metrics per agent and library over each agent's proved cases ($n$).}
    \label{tab:file_read}
    \begin{tabular}{llrrrrrrrr}
        \toprule
        Agent           & Lib     & $n$ & $|\mathit{DR}|$ & $|\mathit{IR}|$ & $|R|$ & $|C|$ & $|{R\cap C}|$ & $\eta$ & $\rho$  \\
        \midrule
        claude-opus-5.5 & aeneas  & 116 & 603             & 990             & 1593  & 744   & 467           & 29.3\% & 62.8\%  \\
        claude-opus-5.5 & mathlib & 116 & 2               & 73              & 75    & 915   & 56            & 74.7\% & 6.1\%   \\
        \addlinespace
        dsh-deepseek    & aeneas  & 17  & 54              & 233             & 287   & 30    & 30            & 10.5\% & 100.0\% \\
        dsh-deepseek    & mathlib & 17  & 17              & 135             & 152   & 81    & 36            & 23.7\% & 44.4\%  \\
        \addlinespace
        kimi-k3         & aeneas  & 10  & 20              & 59              & 79    & 22    & 18            & 22.8\% & 81.8\%  \\
        kimi-k3         & mathlib & 10  & 6               & 54              & 60    & 37    & 12            & 20.0\% & 32.4\%  \\
        \addlinespace
        glm-5.3         & aeneas  & 16  & 88              & 132             & 220   & 27    & 24            & 10.9\% & 88.9\%  \\
        glm-5.3         & mathlib & 16  & 23              & 70              & 93    & 55    & 19            & 20.4\% & 34.5\%  \\
        \addlinespace
        gpt-5.6-sol     & aeneas  & 20  & 107             & 80              & 187   & 130   & 59            & 31.6\% & 45.4\%  \\
        gpt-5.6-sol     & mathlib & 20  & 19              & 130             & 149   & 81    & 16            & 10.7\% & 19.8\%  \\
        \addlinespace
        gpt-6-sol       & aeneas  & 29  & 113             & 176             & 289   & 154   & 105           & 36.3\% & 68.2\%  \\
        gpt-6-sol       & mathlib & 29  & 12              & 162             & 174   & 136   & 38            & 21.8\% & 27.9\%  \\
        \bottomrule
    \end{tabular}
\end{table}

\paragraph{Mathlib knowledge is largely recalled rather than retrieved.}
Opus 5.5 rarely opens Mathlib files, yet its proofs contain many Mathlib citations: only 6.1\% of cited Mathlib files were accessed during the run. This pattern is shared by the other agents, for which 55.6--80.2\% of cited Mathlib files were never accessed. This suggests that agents can often draw on pretrained mathematical knowledge without consulting library source files. In contrast, $\rho$ is consistently higher for Aeneas, indicating a greater reliance on in-context exploration to discover its specialized verification infrastructure.

\paragraph{Aeneas exploration is inefficient for several agents.}
Agents differ substantially in how efficiently they explore Aeneas. GPT-6, GPT-5.6, and Opus 5.5 achieve the highest exploration efficiencies, while DeepSeek V4.1 and GLM 5.3 have particularly low $\eta$. Since Aeneas knowledge is less readily available from pretraining, inefficient exploration can consume a substantial portion of the tool-call budget without contributing to the final proof. These results suggest that helping agents locate relevant Aeneas lemmas could improve their verification performance.

\subsubsection{Ablation: Can Guidance Improve Exploration?}
To test whether reducing exploration costs improves proof success, we provide DeepSeek V4.1 Flash with a case-agnostic guidance file. We choose this agent because it has one of the lowest exploration efficiencies on Aeneas. The guidance is appended to the prompt and made available in the workspace. It contains environment information, a compile-checked prelude of bridge lemmas, loop-proof templates, a map of relevant lemmas, and tool-budget guidelines, but no benchmark testcase names or solutions.

We run the guided agent on the 142 testcases that DeepSeek V4.1 failed to solve originally. The guidance enables it to solve 34 additional testcases, increasing its total number of verified proofs from 17 to 51.

\begin{table}[htbp]
    \centering
    \caption{Guidance ablation on Deepseek V4.1 Flash.}
    \label{tab:ablation}
    \begin{tabular}{lrrrrrrrrrr}
        \toprule
                & array & crypto & dp & ds & graph & integer & matrix & string & tensor & \textbf{Total} \\
        \midrule
        Failed  & 13    & 6      & 20 & 17 & 34    & 2       & 20     & 25     & 5      & \textbf{142}   \\
        Flipped & 6     & 1      & 7  & 4  & 0     & 2       & 3      & 7      & 4      & \textbf{34}    \\
        \bottomrule
    \end{tabular}
\end{table}

We further compare file-reading behaviour on the newly verified cases with that of the original successful cases. Table~\ref{tab:ablation_reads} reports the results.

\begin{table}[htbp]
    \centering
    \caption{Read/use statistics on Comparator-verified proved classifies.}
    \label{tab:ablation_reads}
    \begin{tabular}{llrrrrrr}
        \toprule
        Condition & Lib     & $n$ & $|R|$ & $|C|$ & $\eta$ & $\rho$  & files / case \\
        \midrule
        unguided  & aeneas  & 17  & 287   & 30    & 10.5\% & 100.0\% & 16.9         \\
        unguided  & mathlib & 17  & 152   & 81    & 23.7\% & 44.4\%  & 8.9          \\
        \addlinespace
        guided    & aeneas  & 34  & 407   & 262   & 43.2\% & 67.2\%  & 12.0         \\
        guided    & mathlib & 34  & 310   & 174   & 25.2\% & 44.8\%  & 9.1          \\
        \bottomrule
    \end{tabular}
\end{table}

For Aeneas, guidance increases exploration efficiency $\eta$ from 10.5\% to 43.2\% and reduces the number of accessed files per testcase from 16.9 to 12.0. Mathlib exploration efficiency remains nearly unchanged. Meanwhile, Aeneas pretraining coverage $\rho$ decreases from 100\% to 67.2\%. This is expected: the guidance provides relevant lemma names directly, allowing the agent to use them without first discovering their source files.

The guided proofs also cite substantially more Aeneas files. This suggests that guidance does more than reduce unnecessary exploration; it enables the agent to make use of a broader range of the Aeneas library. Together, the results support our hypothesis that inefficient library exploration limits proof success. However, guidance does not eliminate all difficulties: no graph testcase is solved, suggesting that reasoning about complex program structures remains a separate challenge.

\section{Discussion and Conclusion}
\label{sec:discussion}
We presented \autospec{}, a Rust-to-Lean toolchain and benchmark for studying how coding agents verify existing software implementations. By extending Aeneas with numerical abstraction, \autospec{} connects general-purpose programming with mathematical reasoning in Lean. Its benchmark of 159 testcases across nine algorithmic domains separates implementation from specification proving, enabling a focused evaluation of agents' verification capabilities. We further analyzed proof structure and agent behaviour to identify challenges that can guide the development of future vericoding systems.

Our experiments reveal substantial room for improvement: even the strongest agent verifies only 50.0\% on Tier 2. The analysis identifies two complementary bottlenecks. Agents must acquire verification-specific knowledge from Aeneas and its supporting libraries, while also reasoning about structural proof obligations arising from loops and recursion. Our guidance ablation provides effective solution for the first problem, but the ability to reason about complex program structures remain a bottleneck for future improvement in vericoding systems.

Several limitations remain. The current toolchain supports only a subset of Rust, excluding features such as unsafe code, general mutable references, generic types, and trait-heavy abstractions. Moreover, numerical retargeting interprets selected \texttt{f64} operations over exact rational or real numbers rather than IEEE-754 values. Thus, proofs establish properties of translated programs under these mathematical abstractions, not bit-precise correctness of the original machine-level computations.

Addressing these limitations opens meaningful directions for future research. Broader Rust language support and compositional verification methods could extend translation-based verification to larger software projects. More faithful numerical semantics, including rounding-aware floating-point arithmetic, would also strengthen the connection between mathematical proofs and machine-level execution. Verified floating-point libraries provide a feasible foundation for this direction~\citep{floatlib}.

We hope \autospec{} will serve as a testbed for studying automated program verification and help advance coding agents that can not only produce software, but also establish its correctness.

\bibliographystyle{plainnat}
\bibliography{references}

\newpage

\appendix

\section{LU Decomposition Example}\label{app:lu-example}

This appendix gives the complete Rust implementation, auto-generated Lean translation, and hand-written specification for LU decomposition with partial pivoting
(\texttt{matrix\_lu}) testcase, illustrating the full pipeline described in \S\ref{sec:toolchain}.

\subsection*{Rust implementation}

The function \texttt{lu\_decomposition} takes a  lattened $n \times n$ row-major matrix of \texttt{f64} and returns a packed \texttt{LU} struct. Each column $k$ is processed in three phases: pivot search (lines 17--26), optional row swap (lines 29--38), and Gaussian elimination (lines 41--50).

\begin{lstlisting}[style=code,language=C]
pub struct LU {
    pub lu:   Vec<f64>,   // packed L/U factors, row-major
    pub perm: Vec<usize>, // perm[i] = original row now at row i
}

pub fn lu_decomposition(matrix: &[f64], n: usize) -> LU {
    if n == 0 {
        return LU { lu: Vec::new(), perm: Vec::new() };
    }
    let mut lu = matrix.to_vec();
    let mut perm: Vec<usize> = Vec::new();
    let mut i: usize = 0;
    while i < n { perm.push(i); i += 1; }  // identity permutation

    let mut k: usize = 0;
    while k < n {
        // Partial pivoting: find row with largest |value| in column k
        let mut max_abs = lu[k * n + k];
        if max_abs < 0.0 { max_abs = -max_abs; }
        let mut piv = k;
        let mut i: usize = k + 1;
        while i < n {
            let mut abs_val = lu[i * n + k];
            if abs_val < 0.0 { abs_val = -abs_val; }
            if abs_val > max_abs { max_abs = abs_val; piv = i; }
            i += 1;
        }
        let diag = lu[piv * n + k];
        assert!(diag != 0.0, "singular matrix");

        // Row swap (if needed)
        if piv != k {
            let tmp_perm = perm[k];
            perm[k] = perm[piv]; perm[piv] = tmp_perm;
            let mut j: usize = 0;
            while j < n {
                let tmp = lu[k * n + j];
                lu[k * n + j] = lu[piv * n + j];
                lu[piv * n + j] = tmp;
                j += 1;
            }
        }

        // Gaussian elimination: eliminate rows below the pivot
        let mut i: usize = k + 1;
        while i < n {
            let mult = lu[i * n + k] / diag;
            lu[i * n + k] = mult;      // store multiplier in L slot
            let mut j: usize = k + 1;
            while j < n {
                lu[i * n + j] = lu[i * n + j] - mult * lu[k * n + j];
                j += 1;
            }
            i += 1;
        }
        k += 1;
    }
    LU { lu, perm }
}
\end{lstlisting}

\subsection*{Auto-generated Lean translation}

Aeneas translates each Rust loop into a tail-recursive Lean function annotated with \texttt{@[rust\_loop]}. The seven helpers correspond to: initialising the permutation vector (\texttt{loop0}), the inner pivot search (\texttt{loop1\_loop0}), the row swap (\texttt{loop1\_loop1}), the outer and inner Gaussian elimination (\texttt{loop1\_loop2} and \texttt{loop1\_loop2\_loop0}), and duplicates for the non-swap branch.

The top-level function sequences them via monadic \texttt{do}-notation. The \texttt{f64} constant \texttt{0.0} is inlined by the post-processor as the rational literal \texttt{0}. As stated previously, all arithmetic is over the mathematical \texttt{Rat} domain.

\begin{lstlisting}[style=code]
namespace matrix_lu

structure LU where
  lu   : alloc.vec.Vec Rat
  perm : alloc.vec.Vec Std.Usize

-- Loop 0: initialise permutation vector to identity
@[rust_loop]
def lu_decomposition_loop0
  (n : Std.Usize) (perm : alloc.vec.Vec Std.Usize)
  (i : Std.Usize) :
  Result (alloc.vec.Vec Std.Usize) := do
  if i < n then
    let perm1 <- alloc.vec.Vec.push perm i
    lu_decomposition_loop0 n perm1 (<- i + 1)
  else ok perm
partial_fixpoint

-- Loop 2: inner pivot search (rows k+1 .. n-1 of column k)
@[rust_loop]
def lu_decomposition_loop1_loop0
  (n : Std.Usize) (lu : alloc.vec.Vec Rat) (k : Std.Usize)
  (max_abs : Rat) (piv : Std.Usize) (i : Std.Usize) :
  Result Std.Usize := do
  if i < n then
    let abs_val <-
      alloc.vec.Vec.index (core.slice.index.SliceIndexUsizeSlice Rat)
        lu (<- i * n + k)
    let abs_val1 := if abs_val < 0 then -abs_val else abs_val
    let (max_abs1, piv1) :=
      if abs_val1 > max_abs then (abs_val1, i) else (max_abs, piv)
    lu_decomposition_loop1_loop0 n lu k max_abs1 piv1 (<- i + 1)
  else ok piv
partial_fixpoint

-- Loop 3: row swap (columns 0 .. n-1)
@[rust_loop]
def lu_decomposition_loop1_loop1
  (n : Std.Usize) (lu : alloc.vec.Vec Rat)
  (k piv j : Std.Usize) :
  Result (alloc.vec.Vec Rat) := do
  if j < n then
    let tmp <- alloc.vec.Vec.index
                 (core.slice.index.SliceIndexUsizeSlice Rat)
                 lu (<- k * n + j)
    let fl  <- alloc.vec.Vec.index
                 (core.slice.index.SliceIndexUsizeSlice Rat)
                 lu (<- piv * n + j)
    let (_, back) <- alloc.vec.Vec.index_mut
                       (core.slice.index.SliceIndexUsizeSlice Rat)
                       lu (<- k * n + j)
    let lu1 := back fl
    let (_, back1) <- alloc.vec.Vec.index_mut
                        (core.slice.index.SliceIndexUsizeSlice Rat)
                        lu1 (<- piv * n + j)
    lu_decomposition_loop1_loop1 n (back1 tmp) k piv (<- j + 1)
  else ok lu
partial_fixpoint

-- Loop 5: inner elimination (update row i, columns k+1 .. n-1)
@[rust_loop]
def lu_decomposition_loop1_loop2_loop0
  (n : Std.Usize) (lu : alloc.vec.Vec Rat)
  (k i : Std.Usize) (mult : Rat) (j : Std.Usize) :
  Result (alloc.vec.Vec Rat) := do
  if j < n then
    let fl  <- alloc.vec.Vec.index
                 (core.slice.index.SliceIndexUsizeSlice Rat)
                 lu (<- i * n + j)
    let fl1 <- alloc.vec.Vec.index
                 (core.slice.index.SliceIndexUsizeSlice Rat)
                 lu (<- k * n + j)
    let (_, back) <- alloc.vec.Vec.index_mut
                       (core.slice.index.SliceIndexUsizeSlice Rat)
                       lu (<- i * n + j)
    lu_decomposition_loop1_loop2_loop0
      n (back (fl - mult * fl1)) k i mult (<- j + 1)
  else ok lu
partial_fixpoint

-- Loop 4: outer elimination (iterate rows k+1 .. n-1)
@[rust_loop]
def lu_decomposition_loop1_loop2
  (n : Std.Usize) (lu : alloc.vec.Vec Rat) (k : Std.Usize)
  (diag : Rat) (i : Std.Usize) :
  Result (alloc.vec.Vec Rat) := do
  if i < n then
    let fl   <- alloc.vec.Vec.index
                  (core.slice.index.SliceIndexUsizeSlice Rat)
                  lu (<- i * n + k)
    let mult := fl / diag
    let (_, back) <- alloc.vec.Vec.index_mut
                       (core.slice.index.SliceIndexUsizeSlice Rat)
                       lu (<- i * n + k)
    let lu1 := back mult
    let lu2 <- lu_decomposition_loop1_loop2_loop0
                 n lu1 k i mult (<- k + 1)
    lu_decomposition_loop1_loop2 n lu2 k diag (<- i + 1)
  else ok lu
partial_fixpoint

-- Loop 1: outer loop over pivot column k = 0 .. n-1
@[rust_loop]
def lu_decomposition_loop1
  (n : Std.Usize) (lu : alloc.vec.Vec Rat)
  (perm : alloc.vec.Vec Std.Usize) (k : Std.Usize) :
  Result ((alloc.vec.Vec Rat) * (alloc.vec.Vec Std.Usize)) := do
  if k < n then
    let max_abs <- alloc.vec.Vec.index
                    (core.slice.index.SliceIndexUsizeSlice Rat)
                    lu (<- k * n + k)
    let max_abs1 := if max_abs < 0 then -max_abs else max_abs
    let piv  <- lu_decomposition_loop1_loop0
                  n lu k max_abs1 k (<- k + 1)
    let diag <- alloc.vec.Vec.index
                  (core.slice.index.SliceIndexUsizeSlice Rat)
                  lu (<- piv * n + k)
    massert (diag != 0)
    let (lu1, perm1) <-
      if piv != k then do
        let tp <- alloc.vec.Vec.index
                    (core.slice.index.SliceIndexUsizeSlice Std.Usize) perm k
        let pi <- alloc.vec.Vec.index
                    (core.slice.index.SliceIndexUsizeSlice Std.Usize) perm piv
        let (_, b)  <- alloc.vec.Vec.index_mut
                         (core.slice.index.SliceIndexUsizeSlice Std.Usize)
                         perm k
        let (_, b1) <- alloc.vec.Vec.index_mut
                         (core.slice.index.SliceIndexUsizeSlice Std.Usize)
                         (b pi) piv
        let lu1 <- lu_decomposition_loop1_loop1 n lu k piv 0#usize
        pure (<- lu_decomposition_loop1_loop2
                   n lu1 k diag (<- k + 1), b1 tp)
      else
        pure (<- lu_decomposition_loop1_loop2
                   n lu k diag (<- k + 1), perm)
    lu_decomposition_loop1 n lu1 perm1 (<- k + 1)
  else ok (lu, perm)
partial_fixpoint

def lu_decomposition (matrix : Slice Rat) (n : Std.Usize) :
    Result LU := do
  if n = 0#usize then
    ok { lu   := alloc.vec.Vec.new Rat,
         perm := alloc.vec.Vec.new Std.Usize }
  else
    let lu   <- alloc.slice.Slice.to_vec RatClone matrix
    let perm <- lu_decomposition_loop0
                  n (alloc.vec.Vec.new Std.Usize) 0#usize
    let (lu1, perm1) <- lu_decomposition_loop1 n lu perm 0#usize
    ok { lu := lu1, perm := perm1 }

end matrix_lu
\end{lstlisting}

\subsection*{Lean specification}

The specification bridges the Aeneas output to Mathlib's matrix algebra. Four helper definitions perform the structural translation before the correctness theorem is stated.

\begin{itemize}
  \item \textbf{\texttt{toMatrix}} (lines 3--14) reinterprets a flat \texttt{Slice Rat} as a Mathlib \texttt{Matrix (Fin rows) (Fin cols) Rat} by mapping index pair $(i,j)$ to the element at position $i \cdot \mathit{cols} + j$. The index-bound proof is carried inline so the definition remains computable.
  \item \textbf{\texttt{extractL}} (lines 17--21) and \textbf{\texttt{extractU}} (lines 24--27) split the packed \texttt{LU} matrix into its two triangular factors. The Gaussian elimination stores multipliers in the strict lower triangle and $U$ in the upper triangle. \texttt{extractL} restores the implicit unit diagonal and zeroes the upper triangle. \texttt{extractU} zeroes the strict lower triangle.
  \item  \textbf{\texttt{permMatrix}} (lines 30--38) converts the permutation vector into a $0/1$ permutation matrix: $P_{ij} = 1$ iff $\texttt{perm}[i] = j$.
\end{itemize}

The \textbf{correctness theorem} (lines 41--53) states $P \cdot A = L \cdot U$: the permuted input matrix equals the product of the extracted factors. The length hypotheses \texttt{h\_len}, \texttt{h\_lu\_len}, and \texttt{h\_perm\_len} are needed to instantiate \texttt{toMatrix} and \texttt{permMatrix}.
The hypothesis \texttt{h} binds the result monad: a \texttt{lu\_decomposition matrix n = ok lu} return means the pivot was non-zero at every elimination step, so non-singularity of the input is implicitly proved as a side effect of the successful computation.

\begin{lstlisting}[style=code]
import LeanMode.matrix_lu.MatrixLu
import Mathlib.LinearAlgebra.Matrix.Determinant.Basic
import Mathlib.Tactic

open Aeneas Aeneas.Std Result Matrix matrix_lu

-- (1) Reinterpret a flat Slice Rat as a Mathlib matrix
def toMatrix (s : Slice Rat) (rows cols : Nat)
    (h : s.length = rows * cols) :
    Matrix (Fin rows) (Fin cols) Rat :=
  fun i j =>
    have hb : i.val * cols + j.val < s.length := by
      rw [h]; calc i.val * cols + j.val
          < i.val * cols + cols :=
              Nat.add_lt_add_left j.is_lt _
        _ = (i.val + 1) * cols  := by ring
        _ <= rows * cols         :=
              Nat.mul_le_mul_right cols
                (Nat.succ_le_of_lt i.is_lt)
    s[i.val * cols + j.val]'hb

-- (2) Extract unit lower-triangular factor L from packed matrix
def extractL {n : Nat} (lu : Matrix (Fin n) (Fin n) Rat) :
    Matrix (Fin n) (Fin n) Rat :=
  fun i j => if i.val = j.val then 1
             else if i.val > j.val then lu i j
             else 0

-- (3) Extract upper-triangular factor U from packed matrix
def extractU {n : Nat} (lu : Matrix (Fin n) (Fin n) Rat) :
    Matrix (Fin n) (Fin n) Rat :=
  fun i j => if i.val <= j.val then lu i j else 0

-- (4) Convert permutation vector to 0/1 permutation matrix
def permMatrix (perm_vec : alloc.vec.Vec Std.Usize)
    (h_len : (alloc.vec.Vec.deref perm_vec).val.length = n) :
    Matrix (Fin n) (Fin n) Rat :=
  fun i j =>
    let lst := (alloc.vec.Vec.deref perm_vec).val
    have hb : i.val < lst.length := by rw [h_len]; exact i.is_lt
    let k := (lst.get (Fin.mk i.val hb)).val
    let p : Fin n := if h : k < n then Fin.mk k h else i
    if p = j then 1 else 0

-- (5) Main correctness theorem: P * A = L * U
theorem lu_decomposition_correct
    (matrix : Slice Rat) (n : Std.Usize) (lu : LU)
    (h_len      : matrix.length = n.val * n.val)
    (h          : lu_decomposition matrix n = ok lu)
    (h_lu_len   : (Vec.deref lu.lu).val.length = n.val * n.val)
    (h_perm_len : (Vec.deref lu.perm).val.length = n.val) :
    permMatrix lu.perm h_perm_len *
      toMatrix matrix n.val n.val h_len =
    extractL (toMatrix (Vec.deref lu.lu) n.val n.val h_lu_len) *
    extractU (toMatrix (Vec.deref lu.lu) n.val n.val h_lu_len) := by
  sorry
\end{lstlisting}

\section{Proof Tree Definition}\label{app:proof-tree}

The proof-structure analysis in section \ref{sec:experiment} is computed on proof trees extracted from elaborated Lean proofs. This appendix defines the tree, describes how it is reconstructed from the records of the Lean elaborator, and defines the measurements taken on it.

\subsection*{Tree Structure}

A proof tree represents one Lean theorem or lemma. It alternates between two kinds of nodes:

\begin{itemize}
  \item A \textbf{goal node} holds a proof state: a local context (the hypotheses) and a target (the proposition to prove). Its child is the tactic applied to it. A goal node without a tactic is an open goal: a goal that the tactic block creating it leaves unclosed, to be discharged outside that block.
  \item A \textbf{tactic node} holds one tactic application: its name, its source span, the goals before and after it runs, and the premises it cites. Its children are the goals it produces, so a tactic that closes its goal has no children.
\end{itemize}

Every path from the root alternates between goal and tactic nodes. The root is a goal node holding the theorem statement. A tactic that splits its goal, such as \texttt{constructor} or \texttt{cases}, has several child goals and branches the tree. A tactic that carries its own tactic block, such as \texttt{have h : P := by \ldots}, additionally opens a nested subtree for that block. A theorem whose proof is a term with several embedded tactic blocks has one root per block. A tactic that acts on several goals at once is a child of each of them, so the structure is in general a directed acyclic graph.

A premise is a named constant (a lemma, definition, or instance) referenced by a tactic. It records the constant's fully qualified name and the file that defines it, which determines provenance classes in section \ref{sec:experiment}. A premise that names another theorem of the same spec file is a dependency between theorems, and these dependencies form the theorem-dependency graph used in the organization analysis.

\subsection*{Tree Construction}

The tree is rebuilt from a flat, position-annotated record of every tactic and every constant reference that the Lean elaborator produced for the file.

\begin{enumerate}
  \item \textbf{Extraction.} A Lean meta program traverses the elaborator's information trees. For each tactic, it emits the name, the source span, and the goal lists before and after the tactic. Only tactics that are steps of a tactic sequence and change the proof state are recorded. For each constant reference, it emits the constant's name, its defining file, and the position of the use. A reference located at the constant's own definition is recorded as a declaration instead, together with the source spans of its statement and body.
  \item \textbf{Goal diff.} For each tactic, the goals before and after are compared as multisets of targets. A goal whose target occurs fewer times afterwards is consumed, and one whose target occurs more times is produced. Goals that merely pass through are in neither set.
  \item \textbf{Premise attachment.} Each constant reference is attached to the innermost tactic whose span contains it. This is computed by a sweep over the references in position order, with a heap of the currently open tactic spans ordered by length, so a reference inside a nested tactic block is credited to the inner tactic rather than to its enclosing combinator.
  \item \textbf{Declaration partition.} Each tactic is assigned to the declaration whose name precedes it and whose body contains it, so the tree of each theorem is built independently. Tactics in a declaration's statement, in the bodies of definitions and instances, and in anonymous \texttt{example}s are discarded.
  \item \textbf{Nesting.} Within a theorem, tactics appear in depth-first order of the syntax tree. A tactic $T'$ wraps $T$ if the span of $T'$ strictly contains that of $T$. Assigning each tactic to its innermost wrapper partitions the tactics into subproofs: the top-level subproof, and one subproof per wrapping tactic holding the tactics directly inside it.
  \item \textbf{Linking inside a subproof.} The tactics of a subproof are processed in source order. Each goal consumed by a tactic is linked to the most recent earlier tactic in the subproof that produced it, matching goals on both context and target. A tactic that consumes nothing, such as one that only rewrites a hypothesis, is linked to the preceding tactic through every goal it starts from. A goal that a tactic produced but no later tactic in the subproof consumed becomes an open goal; this covers branches that a combinator leaves to be closed elsewhere.
  \item \textbf{Linking across subproofs.} A wrapping tactic is connected to the entry points of its subproof, i.e., the tactics of the subproof without an incoming link, through the goals those entry points start from. The entry points of the top-level subproof become the roots of the tree.
\end{enumerate}

\begin{figure}[htbp]
  \centering
  \includegraphics[width=0.8\linewidth]{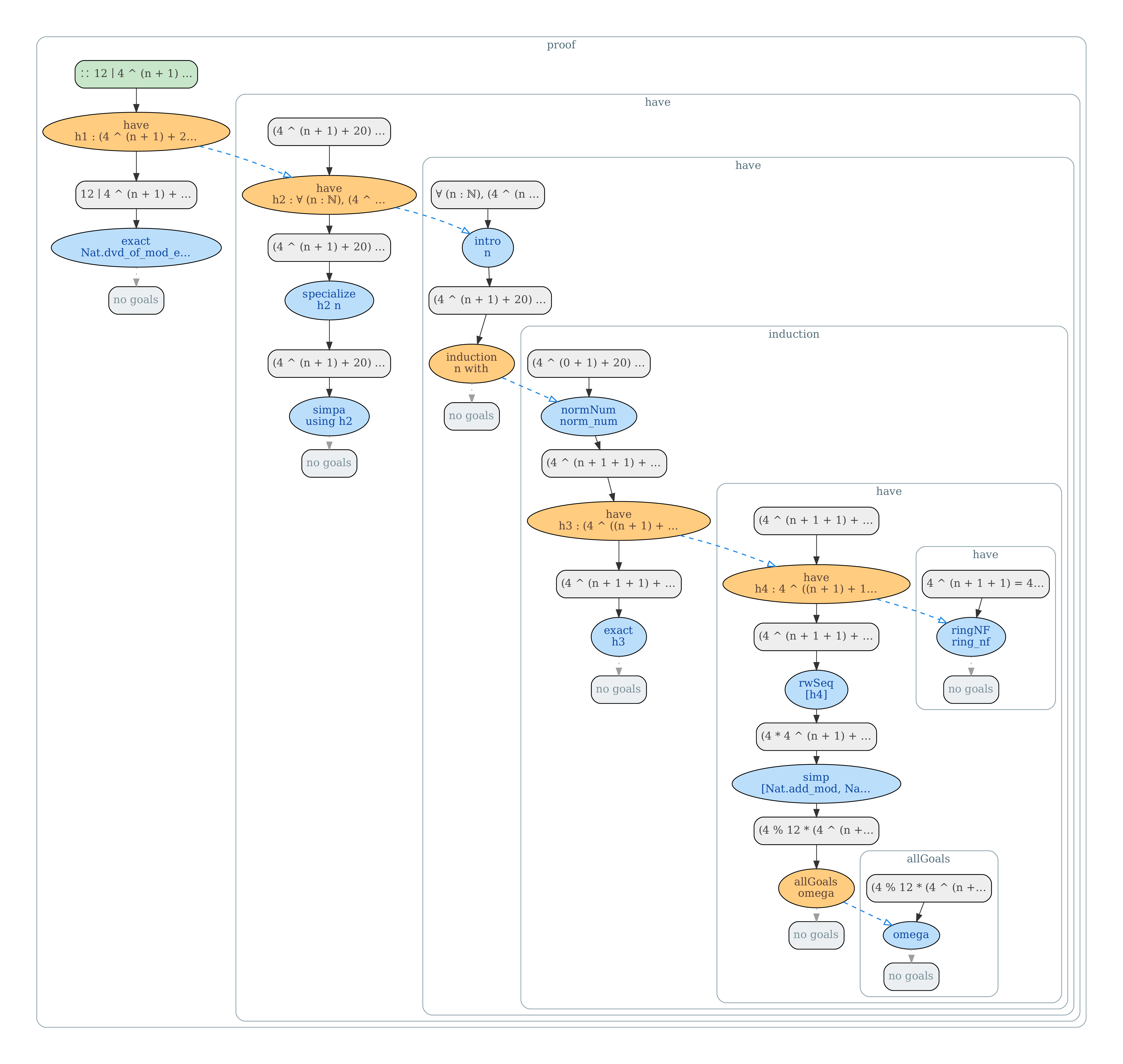}
  \caption{Proof tree of a Lean proof of $12 \mid 4^{n+1} + 20$.}
  \label{fig:proof_tree_example}
\end{figure}

As an example, consider the following Lean proof of $12 \mid 4^{n+1} + 20$:

\begin{lstlisting}[style=code,literate={ℕ}{{$\mathbb{N}$}}1 {∣}{{$\mid$}}1 {∀}{{$\forall$}}1 {⊢}{{$\vdash$}}1]
theorem my_theorem (n : ℕ) : 12 ∣ 4 ^ (n + 1) + 20 := by
  have h1 : (4 ^ (n + 1) + 20) % 12 = 0 := by
    have h2 : ∀ (n : ℕ), (4 ^ (n + 1) + 20) % 12 = 0 := by
      intro n
      induction n with
      | zero =>
        norm_num
      | succ n ih =>
        have h3 : (4 ^ ((n + 1) + 1) + 20) % 12 = 0 := by
          have h4 : 4 ^ ((n + 1) + 1) = 4 * (4 ^ (n + 1)) := by
            ring_nf
          rw [h4]
          simp [Nat.add_mod, Nat.mul_mod] at ih ⊢
          all_goals
            omega
        exact h3
    specialize h2 n
    simpa using h2
  exact Nat.dvd_of_mod_eq_zero h1
\end{lstlisting}

Figure~\ref{fig:proof_tree_example} shows its proof tree. The proof first establishes the remainder claim \texttt{h1}, deriving it from the general statement \texttt{h2} proved by induction, and then closes the goal with \texttt{exact}. A solid arrow shows a goal flowing into the tactic applied to it, and out of that tactic into the goals it produces. Composite tactics such as \texttt{have}, \texttt{induction}, and \texttt{all\_goals} contain a nested subproof, drawn as an enclosing box and attached to the tactic by a dashed blue arrow; primitive tactics such as \texttt{rw} and \texttt{omega} have none. Nesting is thus how the tree represents intermediate claims, case branches, and combinators.

\subsection*{Measurements}

All statistics of section \ref{sec:experiment} are computed per theorem on this tree.

\begin{itemize}
  \item \textbf{Tactics.} The number of distinct tactic nodes; a node shared by several goals is counted once.
  \item \textbf{Depth.} The number of tactic steps on the longest path from the root to a goal node. A proof that proceeds strictly step by step has depth close to its tactic count.
  \item \textbf{Premises.} The number of distinct premises cited by the tactics of the theorem. For a testcase, it is the number of distinct premise names over all of its theorems.
  \item \textbf{Provenance.} Each premise citation, i.e., a premise cited by a tactic node, is classified by the file that defines the constant, and each class is reported as a share of all citations.
  \item \textbf{Tactic buckets.} Each tactic node is assigned by name to one of the eight buckets of Table~\ref{tab:tactic_buckets}. Tactics that the extractor reports under internal syntax-node names (e.g.\ \texttt{tacticHave\_\_} for \texttt{have}) are assigned to the bucket of the tactic they implement.
\end{itemize}

The tree has two limitations. First, only tactic-mode proofs are decomposed: constants used inside a term embedded in a tactic count as premises of that tactic, but the term itself is not broken into nodes, and a proof written entirely in term mode has no tree. Second, links between tactics are inferred by matching pretty-printed goals, so two syntactically identical goals in one subproof are told apart only by their source order.

\section{Reading Behaviour: Phases and Transitions}\label{app:hmm}

This appendix examines the agent-specific transition patterns underlying the reading behaviour analysis in Section~\ref{sec:hmm}.

\paragraph{Setup.}
We map each tool call to one of five symbols: \texttt{read} (file-reading tools and read-only shell commands), \texttt{write} (file edits), \texttt{build} (\texttt{lake} and \texttt{lean} invocations), \texttt{mcp} (queries to the Lean language server), and \texttt{other} (all remaining calls). We fit a five-state categorical HMM to the pooled traces of all eight agents, comprising 114{,}662 calls. The five states are named after their dominant emission symbols. We keep the emission matrix fixed (Table~\ref{tab:hmm_emission}) and fit a separate transition matrix for each agent.

Five states is the largest model size that remains stable across random restarts. Additional states primarily split the language-server phases. Moreover, agent-specific transition matrices fit the traces significantly better than a shared transition matrix. This indicates that agents differ not only in their phase occupancy, but also in how they move between phases.

\begin{table}[htbp]
    \centering
    \caption{Emission probabilities of the fitted HMM (\%).}
    \label{tab:hmm_emission}
    \begin{tabular}{lrrrrr}
        \toprule
        Phase      & read & write & build & mcp  & other \\
        \midrule
        read       & 94.4 & $<$1  & 1.9   & 1.9  & 1.2   \\
        write      & 11.1 & 84.2  & $<$1  & 3.7  & $<$1  \\
        write+read & 22.3 & 51.3  & 17.3  & $<$1 & 8.1   \\
        mcp        & 3.5  & $<$1  & $<$1  & 94.7 & $<$1  \\
        mcp+read   & 14.6 & 3.2   & $<$1  & 81.6 & $<$1  \\
        \bottomrule
    \end{tabular}
\end{table}

\begin{figure}[htbp]
    \centering
    \includegraphics[width=0.8\linewidth]{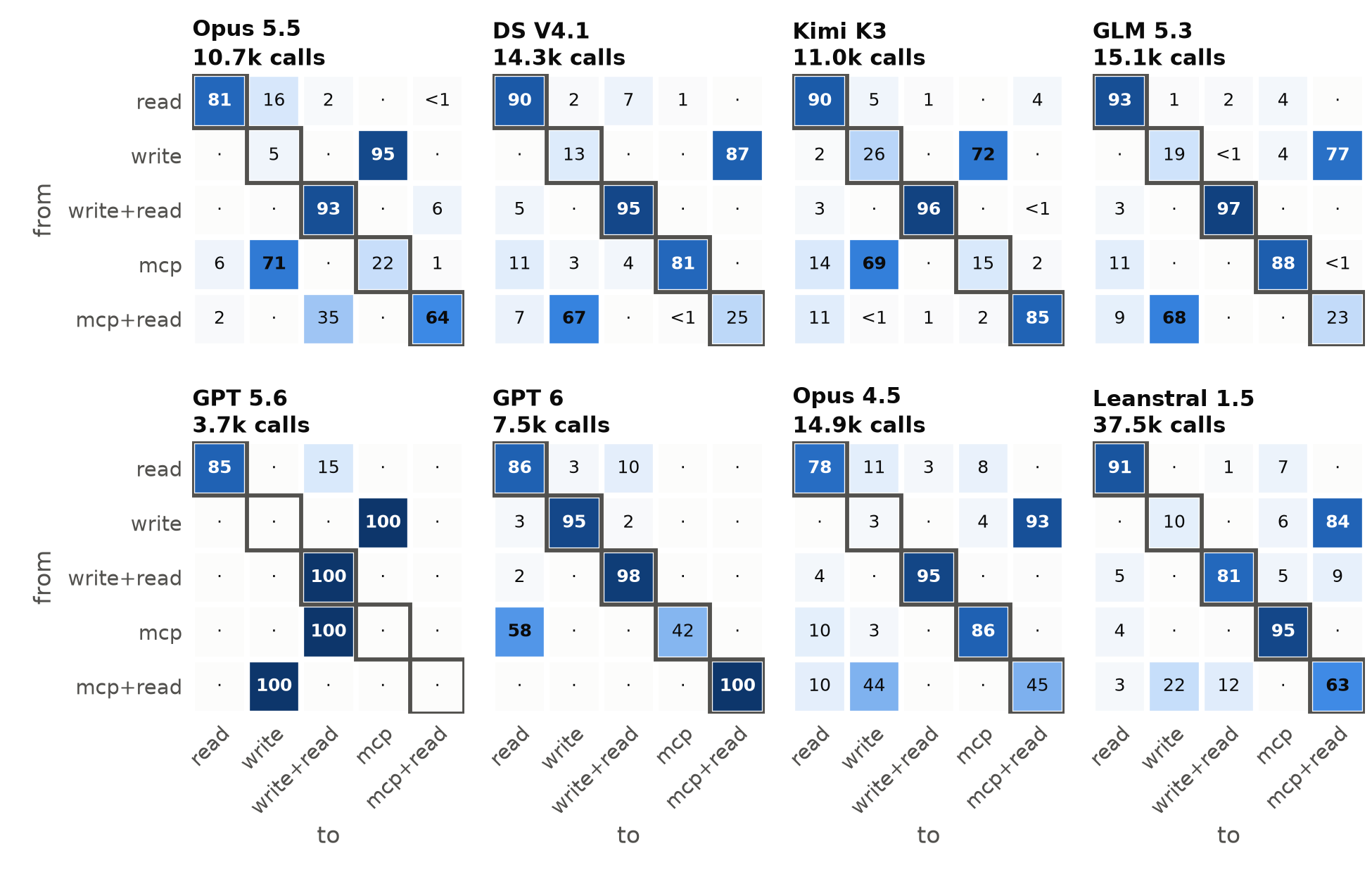}
    \caption{Agent-specific transition matrices (\%).}
    \label{fig:hmm_transitions}
\end{figure}

\paragraph{Interpreting the phases.}
Figure~\ref{fig:hmm_transitions} shows the transition matrix for each agent. The diagonal entry $a_{ii}$ gives the probability of remaining in phase $i$. Accordingly, $1/(1-a_{ii})$ is the expected number of consecutive calls in that phase, or its \emph{dwell time}. Off-diagonal entries describe transitions between phases.

Two drafting phases emerge. The \texttt{write} phase is brief, with a dwell time of approximately one call for agents that use it. In contrast, \texttt{write+read} is much longer, with dwell times of 15--31 calls for most agents. This suggests two distinct drafting patterns: isolated edits followed by checks, and extended drafting episodes interleaved with reads and builds. Across these patterns, the agents exhibit four characteristic workflows.

\paragraph{Claude: edit--check loops with reading up front.}
Both Claude models frequently check edits through the language server. After a \texttt{write} phase, Opus 5.5 transitions to \texttt{mcp} in 100\% of cases, while Opus 4.5 transitions to \texttt{mcp+read} in 96\%. They then return to editing from the corresponding language-server phases in 91\% and 81\% of cases, respectively. Opus 5.5 also exhibits a longer loop between \texttt{write+read} and \texttt{mcp+read}: 89\% of exits from the former lead to the latter, and 96\% of exits from the latter return to drafting. Its reading is concentrated near the beginning of the trace: the read phase has an average dwell time of only 5.4 calls, and fewer than 10\% of transitions from other phases lead back to reading. Opus 4.5 follows a similar edit--check pattern but returns to reading more often: 80\% of exits from \texttt{write+read} and 71\% of exits from \texttt{mcp} lead to \texttt{read}. Thus, Opus 5.5 tends to gather information early and then work against the language server, whereas Opus 4.5 revisits source files more frequently.

\paragraph{DeepSeek, Kimi, and GLM: alternating reading and drafting.}
DeepSeek V4.1, Kimi K3, and GLM 5.3 spend extended periods in both \texttt{read} (10--15 calls) and \texttt{write+read} (20--31 calls), frequently alternating between them. Between 80\% and 95\% of transitions out of \texttt{write+read} lead back to reading. After each drafting episode, these agents therefore tend to revisit the library rather than directly checking their work through the language server. Although short \texttt{write} phases are usually followed by language-server queries (95--100\% of exits), they account for only 4--13\% of these agents' calls. GLM 5.3 exhibits the strongest tendency toward repeated exploration. Only 43\% of its transitions out of \texttt{read} lead to drafting, while 56\% lead to \texttt{mcp}. Of the latter transitions, 90\% return to reading. This creates a read--query loop in which the agent repeatedly explores files and queries the language server, with relatively little transition into proof drafting.

\paragraph{GPT: extended drafting with compilation.}
The two GPT agents primarily alternate between \texttt{read} and \texttt{write+read}. Nearly all transitions out of reading lead to drafting, and almost all transitions out of drafting return to reading. Their drafting episodes are particularly long: the dwell time is 48 calls for GPT-6 and over 400 calls for GPT-5.6. The \texttt{write+read} phase is also the only phase with a substantial share of build calls (17\%). This suggests that the GPT agents primarily check their proofs by invoking \texttt{lake} or \texttt{lean} within extended drafting episodes, rather than relying on frequent language-server queries.

\paragraph{Leanstral: a read--query loop.}
Leanstral 1.5 spends 89\% of its calls in the \texttt{read} and \texttt{mcp} phases, frequently alternating between them: 80\% of transitions out of reading lead to \texttt{mcp}, and 83\% of transitions out of \texttt{mcp} return to reading. Its average dwell time in \texttt{mcp} is approximately 20 calls. Leanstral rarely enters a drafting phase, and its \texttt{write+read} episodes are short, lasting only 5.3 calls on average, compared with at least 15 calls for the other agents. Its trace is therefore dominated by file exploration and language-server queries, with relatively little proof construction. This pattern is consistent with its low accuracy, despite making the most tool calls of any agent (37.5k).

Overall, the transition matrices reveal distinct workflows across agents. Opus 5.5 reads briefly and then works through repeated edit--check cycles. Opus 4.5 follows a similar pattern but revisits source files more frequently. DeepSeek, Kimi, and GLM repeatedly alternate between reading and drafting; the GPT agents use much longer drafting episodes with compilation-based feedback; and Leanstral spends most of its time reading and querying the language server. These patterns suggest that success depends not simply on frequent feedback or extensive exploration, but on how effectively agents turn retrieved information into proof construction. Opus 5.5 stands out for concentrating its reading early and transitioning quickly into drafting.

\end{document}